# CAUSALITY IN LOCAL QUANTUM FIELD THEORY*

**Luciano Maiani**
and
**Massimo Testa**

*Dipartimento di Fisica, Università di Roma "La Sapienza"*
*Sezione INFN di Roma*
*P.le A.Moro 2, 00185 Roma, Italy*

**Abstract**

The problem of causality is analyzed in the context of Local Quantum Field Theory. Contrary to recent claims, it is shown that apparent noncausal behaviour is due to a lack of the the notion of sharp localizability for a relativistic quantum system.

* Partially supported by the EEC grant # SCI-0430-C and by Ministero dell' Università e della Ricerca Scientifica e Tecnologica

In an old paper, Fermi[1] discussed the problem of causality in Quantum Field Theory through a "gedanken" experiment in which he studied the effects induced by the decay of an excited atom *A* on another atom *B* placed at a certain distance *R* (for a review of the older papers see ref.[2]).

Common sense causality suggests that no influence can be transmitted from *A* to *B* before a time $R/c$ has elapsed. This was, in fact, the conclusion reached by Fermi through a perturbative computation. Many years after, this result was shown by M.I.Shirokov[3] to be affected by an unjustified approximation. The observation reopened the question of the causal behaviour of a quantum system, with contrasting conclusions being reached by different authors[4-8]. Very recently the issue has been re-examined in the framework of Local Quantum Field Theory (LQFT) and doubts have been cast on the causal behaviour[9].

The origin of this confusing situation may be traced back to the fact that the notion of localization in the quantum relativistic context is rather involved[10]. This is formalized in Relativistic Field Theory by the so called Reeh-Schlieder Theorem[11], which says that every state can be approximated as closely as one wants by states obtained by applying to the vacuum field polynomials averaged on *any* finite region of space-time: sharp localization is impossible in LQFT.

In view of this situation, and of the fact that experiments of the kind discussed by Fermi are not "gedanken" any more[12], we try to clarify the issue by studying a simple system which can be considered as the prototype of a causality-testing experiment. In view of the previously mentioned difficulties in defining a localized state, we will rather define localization in dynamical terms, by perturbing our system through an action which is localized (in the usual sense) in a given, limited region of space-time. Our conclusion is that LQFT fulfills common sense causality, except for effects which vanish exponentially with the distance and are due to the lack of localizability inherent to relativistic quantum systems[1].

We are fully aware that the arguments presented in the following might be already known to many physicists, although we could not find them stated in the literature in a systematic form.

The physical situation we have in mind is one in which an antenna radiates relativistic quanta associated to a local field $\Phi(x)$, scalar for simplicity. $\Phi(x)$ obeys the fundamental requirement of microcausality, i.e. it commutes with itself at space-like separations:

---

[1] These effects have been found by several authors, for example they correspond to those found in the non-resonant part of the transition probability, in the calculation of ref.[6].

$$[\Phi(x), \Phi(y)]_{(x-y)^2<0} = 0 \qquad (1)$$

The antenna is switched on in a limited space-time region, $O_1$. This situation is described by adding, to the unperturbed action of the field, $S_\Phi$, an interaction term of the form:

$$S_1 = \int_{-\infty}^{+\infty} d^4x\, \lambda_{O_1}(x) \Phi(x) \qquad (2)$$

where $\lambda_{O_1}(x)$ is a given function whith support in $O_1$.

Already at this level one proves[2] that the average value of any local observable $L(y)$ at a point $y$ which is space-like with respect to $O_1$ ($y \sim O_1$) is not influenced by the presence of the antenna[2]. In fact, taking the system in a given initial state $|A\rangle_\Phi$, before the antenna is switched on, we have for its time evolution (in the interaction picture):

$$|A,t\rangle_\Phi = U_{O_1}^{(\Phi)}(t)|A\rangle_\Phi \qquad (3)$$

where:

$$U_{O_1}^{(\Phi)}(t) = T\exp\left( i\int_{-\infty}^{t} d^4x\, \lambda_{O_1}(x)\Phi(x) \right) \qquad (4)$$

The expectation value of $L(y)$ in such a state is:

$$_\Phi\langle A,t|L(\underline{y},t)|A,t\rangle_\Phi = {}_\Phi\langle A|U_{O_1}^{(\Phi)+}(t)L(\underline{y},t)U_{O_1}^{(\Phi)}(t)|A\rangle_\Phi = {}_\Phi\langle A|L(\underline{y},t)|A\rangle_\Phi \qquad (5)$$

and is independent upon the function $\lambda_{O_1}(x)$.

Next we consider the case in which a receiving apparatus, $A$, is operating in another region $O_2$ of space-time. The receiving apparatus can be quite general. It can be schematized as an atom whose coupling with the radiation field $\Phi(x)$ is switched on in the region $O_2$ only. Its presence contributes to the action a term of the form:

$$S_A + S_2 \qquad (6)$$

where $S_A$ is the action which describes the dynamics of $A$, except its interaction with $\Phi(x)$, and:

$$S_2 = \int_{-\infty}^{+\infty} d^4x \, \lambda_{O_2}(x) J_A(x) \Phi(x) \qquad (7)$$

where $J_A(x)$ is the "current" describing the coupling of $A$ to the field $\Phi$.

A few comments about the meaning of eq.(7). The switching function $\lambda_{O_2}(x)$ is introduced in order to allow only the portion of the electrons present inside the region $O_2$ to interact with the radiation. This condition is not realized in a physical atom, of course, whose energy- eigenstates have wave-functions which extend up to spatial infinity. It is clear that if we allow the electron wave-function to be different from zero up to the location of the antenna, we will get a non-vanishing, exponentially small, transition probability. This effect, obviously, does not represent a violation of causality and is completely suppressed with the use the action in eq.(7).

In the case of a nonrelativistic electron, we could start with a sharply localized wave function, but, after any small interval of time, the probability to find the electron in any point of space is different from zero, due to the infinite speed of propagation implied by the diffusion-like (parabolic) character of the nonrelativistic Schrœdinger equation. In a fully relativistic treatment, sharp localization is impossible, due to the Reeh-Schlieder theorem. The spatial cut-off is used in eq.(7) in order, again, not to mix genuine acausal effects with exponentially small non-localization effects.

The evolution of the coupled system "antenna+$\Phi$+A(tom)" is best described in the Dirac picture related to the "interaction" $S_I = S_1 + S_2$. We can therefore write the unitary evolution operator $U(t)$ as:

$$U(t) = T \exp\left( i \int_{-\infty}^{t} d^4x \, \lambda_{O_1}(x) \Phi(x) + i \int_{-\infty}^{t} d^4x \, \lambda_{O_2}(x) J_A(x) \Phi(x) \right) \qquad (8)$$

We will be interested in the case in which the two regions $O_1$ and $O_2$ are space like separated from each other. In this case $U(t)$, due to eq.(1), factorizes into two mutually commuting, unitary operators:

$$U(t) = U_{O_1}^{(\Phi)}(t) U_{O_2}^{(\Phi,A)}(t) \qquad (9)$$

$$\begin{aligned} U_{O_1}^{(\Phi)}(t) &\equiv T \exp\left( i \int_{-\infty}^{t} d^4x \, \lambda_{O_1}(x) \Phi(x) \right) \\ U_{O_2}^{(\Phi,A)}(t) &\equiv T \exp\left( i \int_{-\infty}^{t} d^4x \, \lambda_{O_2}(x) J_A(x) \Phi(x) \right) \end{aligned} \qquad (10)$$

Let us study the consequences of eq.(9). Suppose the state of the system has been prepared, at times antecedent to both regions $O_1$ and $O_2$ as, for example:

$$|i\rangle \equiv |0\rangle_\Phi |G\rangle_A \tag{11}$$

that is a state in which no quanta of the radiation field are present and the atom $A$ is in the ground state. The time evolution of such a state is described by:

$$|i,t\rangle = U_{O_1}^{(\Phi)}(t) U_{O_2}^{(\Phi,A)}(t) |0\rangle_\Phi |G\rangle_A \tag{12}$$

Let us now perform a (non local) measurement at time $t$ and ask for the probability $P_{n,E}(t)$ of detecting the radiation in the state $|n\rangle_\Phi$ and the "atom" in the excited state $|E\rangle_A$. We have:

$$P_{n,E}(t) = $$
$$= {}_A\langle G|{}_\Phi\langle 0| U_{O_2}^{(\Phi,A)+}(t) U_{O_1}^{(\Phi)+}(t) |n\rangle_\Phi |E\rangle_A {}_A\langle E|{}_\Phi\langle n| U_{O_1}^{(\Phi)}(t) U_{O_2}^{(\Phi,A)}(t) |0\rangle_\Phi |G\rangle_A \tag{13}$$

$P_{n,E}(t)$ clearly exhibits correlations between the two regions $O_1$ and $O_2$. However, if we limit ourselves to a measurement on the state of $A$ only, a truly local observation, we have, for the probability $P_E(t)$ of finding $A$ in the excited state $|E\rangle_A$, the expression:

$$P_E(t) = \sum_n P_{n,E}(t) = $$
$$= {}_A\langle G|{}_\Phi\langle 0| U_{O_2}^{(\Phi,A)+}(t) U_{O_1}^{(\Phi)+}(t) |E\rangle_A I_\Phi {}_A\langle E| U_{O_1}^{(\Phi)}(t) U_{O_2}^{(\Phi,A)}(t) |0\rangle_\Phi |G\rangle_A \tag{14}$$

where the completeness for the radiation field has been used and $I_\Phi$ denotes the identity operator in the $\Phi$-Hilbert space. Since the unitary operator $U_{O_1}^{(\Phi)}(t)$ does not act on the $A$ degrees of freedom, eq.(14) becomes:

$$P_E(t) = {}_A\langle G|{}_\Phi\langle 0| U_{O_2}^{(\Phi,A)+}(t) |E\rangle_A {}_A\langle E| U_{O_2}^{(\Phi,A)}(t) |0\rangle_\Phi |G\rangle_A \tag{15}$$

Equation (15) is precisely the statement that whatever happens in the region $O_1$ cannot influence the behaviour of the measuring apparatus, active in the space-like separated region $O_2$.

The above considerations are, in our opinion, exhaustive. However, it can be useful to elaborate a little further, also in order to make a closer contact with the different points of view expressed in the literature.

First of all, we can easily relax the approximation in which the atom $A$ interacts with the radiation $\Phi$ only for a limited time interval. In fact we can consider, instead of $S_2$, an interaction described by $S'_2$:

$$S'_2 = \int_{-\infty}^{+\infty} d^4x\, \lambda_R(\underline{x}) J_A(\underline{x},t) \Phi(\underline{x},t) \qquad (16)$$

in which the "coupling constant" $\lambda_R(\underline{x})$ is time independent and has a spatial support $\underline{R}$ such that its extrusion[2] in space-time does not intersect $O_2$. In this way we switch on the coupling of $A$ with radiation at $t = -\infty$ and allow for a fully "dressed" atom at finite times. Denoting by $t_{O_1}$ the time at which the antenna starts radiating, we have that, at time $t_{O_1}$, the state of the system is:

$$\left|t_{O_1}\right\rangle \equiv T\exp\left(i\int_{-\infty}^{t_{O_1}} d^4x\, \lambda(\underline{x}) J_A(\underline{x},t) \Phi(\underline{x},t)\right) |0\rangle_\Phi |G\rangle_A \qquad (17)$$

The evolution for times subsequent to $t_{O_1}$ is, therefore, described by:

$$\left|t_{O_1},t\right\rangle = U(t,t_{O_1})\left|t_{O_1}\right\rangle \qquad (18)$$

where:

$$U(t,t_{O_1}) = T\exp\left(i\int_{t_{O_1}}^{t} d^4x\, \lambda_{O_1}(x) \Phi(x) + i\int_{t_{O_1}}^{t} d^4x\, \lambda_R(\underline{x}) J_A(\underline{x},t) \Phi(\underline{x},t)\right) \qquad (19)$$

We have again:

$$U(t,t_{O_1}) = U^{(\Phi)}_{O_1}(t,t_{O_1}) U^{(\Phi,A)}_{\lambda_R}(t - t_{O_1}) \qquad (20)$$

$$U^{(\Phi)}_{O_1}(t,t_{O_1}) = T\exp\left(i\int_{t_{O_1}}^{t} d^4x\, \lambda_{O_1}(x) \Phi(x)\right)$$

$$U^{(\Phi,A)}_{\lambda_R}(t - t_{O_1}) = T\exp\left(i\int_{t_{O_1}}^{t} d^4x\, \lambda_R(\underline{x}) J_A(\underline{x},t) \Phi(\underline{x},t)\right) \qquad (21)$$

---

[2] i.e. the cylinder in space-time with basis $\underline{R}$.

for times $t$ such that the space-time region $\underline{R} \otimes (t, t_{O_1})$ is space-like with respect to $O_1$. Also in this case the same argument as the one in eqs.(14), (15), shows that no violations of causality can be attained.

The last physical sistem we want to discuss, is one in which the transmitting antenna is replaced by another atom. This amounts to replace the action given in eq.(2) by:

$$S'_1 = \int_{-\infty}^{+\infty} d^4x\, \lambda_{O_1}(x) J_A(x) \Phi(x) \qquad (22)$$

Since we are using the same atomic "current" $J_A(x)$ for both atomic systems, this choice of the action implies total (anti-) symmetrization of the wave function of the two atoms. In fact this is a subtle case. The "atomic current" $J_A(x)$ satisfies a microcausality condition:

$$\left[J_A(x), J_A(y)\right]_{(x-y)^2 < 0} = 0 \qquad (22)$$

Eq.(22) guarantees the factorization of the evolution matrix as before. However the electrons of the atom localized in $O_1$, when inside the region $O_2$, are allowed to interact with the radiation field. This effect, together with (anti-) symmetrization of the wave function, gives rise to exponentially small correlations which could look like violations of causality. Rather, these effects owe their existence to the fact that the observation of the excited atomic state in $O_1$ requires a non local measurement which depends from the state of affairs existing in $O_2$. In particular, in eq.(15) one is not allowed to commute the evolution operator in $O_2$ with the projection operator over the state $|E\rangle_A$.

This statement is confirmed by considering the hypothetical case of two distinguishable atoms, whose states belong to two different Hilbert spaces[3]. We have:

$$\left[J_A(x), J_B(y)\right] = 0 \qquad (23)$$

identically, because the two currents are built up from two independent degrees of freedom and, in the interaction picture we are adopting, they evolve in time through the "free" action:

$$S_A + S_B \qquad (24)$$

---
[3] In a quantum world without Weak Interactions, we could consider, e.g., e-P and µ-Σ⁺ atoms.

In this case spurious violations of causality are not present.

Let us end with some general remarks.

First of all, radiative corrections usually require regularization and renormalization, in order to give finite results. The need of renormalization does not spoil our general conclusions. In fact in order to renormalize the theory it is enough to add **local** counterterms to the foregoing actions and this operation does not alter the microcausality property, eq.(1), on which the whole argument is based.

A second, more subtle point, is that we have computed the probability of getting, as a final state after the measurement, a "bare" atomic state and not a true eigenstate of the total hamiltonian, including the interaction with the radiation field $\Phi(x)$. For a measurement designed to cause the collapse of the wave function on a true eigenstate of the total hamiltonian non local effects are indeed expected because such an experimental apparatus should detect (exponentially small) effects of the virtual radiation and would correspond to a non local perturbation on the whole quantum system.

We acknowledge useful discussions with Professors F. De Martini and P. Mataloni.